\documentclass[twocolumn]{article}

\usepackage[utf8]{inputenc}
\usepackage{amsfonts,amssymb,amsmath}
\usepackage{graphicx}
\usepackage[authoryear]{natbib}
\usepackage{url}

\title{Reanalysis of the spin direction distribution of Galaxy Zoo SDSS spiral galaxies}

\author{Darius McAdam,  Lior Shamir \\ Kansas State University \\ Manhattan, KS, 66506, USA \\ E-mail: lshamir@mtu.edu}
\date{}

\begin{document}

\maketitle

\begin{abstract}
The distribution of the spin directions of spiral galaxies in the Sloan Digital Sky Survey has been a topic of debate in the past two decades, with conflicting conclusions reported even in cases where the same data was used. Here we follow one of the previous experiments by applying the {\it SpArcFiRe} algorithm to annotate the spin directions in original dataset of Galaxy Zoo 1. The annotation of the galaxy spin directions is done after a first step of selecting the spiral galaxies in three different manners: manual analysis by Galaxy Zoo classifications, by a model-driven computer analysis, and with no selection of spiral galaxies. The results show that when spiral galaxies are selected by Galaxy Zoo volunteers, the distribution of their spin directions as determined by {\it SpArcFiRe} is not random, which agrees with previous reports. When selecting the spiral galaxies using a model-driven computer analysis or without selecting the spiral galaxies at all, the distribution is also not random. Simple binomial distribution analysis shows that the probability of the parity violation to occur by chance is lower than 0.01. Fitting the spin directions as observed from Earth to cosine dependence exhibits a dipole axis with statistical strength of 2.33$\sigma$ to 3.97$\sigma$. These experiments show that regardless of the selection mechanism and the analysis method, all experiments show similar conclusions. These results are aligned with previous reports using other methods and telescopes, suggesting that the spin directions of spiral galaxies as observed from Earth exhibit a dipole axis formed by their spin directions. Possible explanations can be related to the large-scale structure of the Universe, or to internal structure of galaxies. The catalogs of annotated galaxies generated as part of this study are available.
\end{abstract}

%\begin{keywords}
%galaxies: general -- galaxies: spiral -- Cosmology: large-scale structure of universe -- Cosmology: Cosmic anisotropy
%\end{keywords}

%\keywords{Galaxy: general -- galaxies: spiral -- large-scale structure of universe}

\section{Introduction}
\label{introduction}

The distribution of spin directions of spiral galaxies has been a topic of discussion for several decades, with several studies showing conflicting results regarding the same data. In particular, several different studies have shown conflicting results when using galaxies from SDSS, and specifically the set of SDSS galaxies with spectra. Namely, some studies suggest that the distribution of spin direction of SDSS spiral galaxies with spectra is random, while other experiments showed statistically significant asymmetry in the same data. If the distribution of galaxy spin directions does not conform with the parity assumption, the parity violation can be exhibited through a cosmological-scale dipole axis formed by the large-scale distribution of galaxy spin directions.

Several previous studies showed non-random distribution of the spin directions of spiral galaxies in SDSS, and suggested that the distribution forms a large-scale dipole axis \citep{longo2011detection,shamir2012handedness,shamir2019large,shamir2020patterns}. In fact, claims for non-random distribution of galaxy spin directions were reported nearly two decades before SDSS saw first light \citep{macgillivray1985anisotropy}. The non-randomness also showed a statistically significant number of galaxies spinning in opposite directions in opposite hemispheres \citep{shamir2019large,shamir2020patterns,shamir2021large}. That is, the SDSS galaxies can be separated into two hemispheres such that one hemisphere has a higher number of galaxies spinning clockwise, while the opposite hemisphere has a higher number of galaxies spinning counterclockwise, and the differences are statistically significant \citep{shamir2019large,shamir2020patterns,shamir2021large}. In particular, among all SDSS galaxies with spectra, there is a higher number of galaxies spinning counterclockwise compared to the number of galaxies spinning clockwise \citep{longo2011detection,shamir2019large,shamir2020patterns}. Other possibly related studies used a smaller number of galaxies to show alignment in the spin directions of SDSS galaxies that are too far from each other to interact gravitationally \citep{lee2019galaxy,lee2019mysterious,motloch2021observed}. These observations were defined as ``mysterious'', suggesting that the large-scale structure is linked through galaxy spin directions \citep{lee2019mysterious}.

In addition to SDSS, other telescopes also showed parity violation in the distribution of the spin direction of spiral galaxies. These include Pan-STARRS \citep{shamir2020patterns}, DECam \citep{shamir2021large}, Hubble Space Telescope \citep{shamir2020pasa}, the Dark Energy Survey \citep{shamir2022asymmetry}, and the DESI Legacy Survey \citep{shamir2022analysis}. These telescopes cover both the Northern and Southern hemispheres, and the Hubble Space Telescope provides analysis that is not subjected to a possible effect of the atmosphere. On the other hand, other experiments showed no statistically significant difference between the number of clockwise and counterclockwise galaxies \citep{iye1991catalog}, and suggested that the distribution of spin directions in SDSS is random \citep{land2008galaxy,hayes2017nature,iye2020spin}.

However, it is difficult to fully prove the absence of non-random distribution. Experiments that showed randomness in the distribution of galaxy spin directions might not necessarily prove that the distribution of galaxy spin directions is indeed random, but merely that the signal is not statistically significant. Reasons can include datasets that are not sufficiently large, or biases in the data that their corrections require modifications to the data. These reasons are summarized in Section~\ref{previous_work}, and a more detailed discussion can be found in \citep{shamir2022using,shamir2022analysis2,shamir2022asymmetry}.

While the assumption that the Universe is isotropic is part of the Cosmological Principle, several different probes have shown non-random distribution at cosmological scales \citep{aluri2022observable}. These probes include the CMB radiation \citep{eriksen2004asymmetries,cline2003does,gordon2004low,campanelli2007cosmic,copi2010large,zhe2015quadrupole,ashtekar2021cosmic,yeung2022directional,greco2022cosmic}, as well as other probes such as LX-T scaling \citep{migkas2020probing}, cosmic rays \citep{aab2017observation}, short gamma ray bursts \citep{meszaros2019oppositeness}, Ia supernova \citep{javanmardi2015probing,lin2016significance}, dark energy \citep{adhav2011kantowski,adhav2011lrs,perivolaropoulos2014large,colin2019evidence}, quasars \citep{quasars,zhao2021tomographic,secrest2021test},  $H_o$ \citep{luongo2022larger,dainotti2022evolution}, and galaxy shapes \citep{javanmardi2017anisotropy}. It has also been shown that the large-scale structure might have ``handedness``, exhibited by asymmetry of the four-point correlation function such that each point is a galaxy \citep{philcox2022probing,hou2022measurement}.

If these observations reflect the real large-scale structure of the Universe, they shift from the Cosmological Principle and the standard cosmological models \citep{aluri2022observable}, and can be related to several alternative theories. These include ellipsodial Universe \citep{campanelli2006ellipsoidal,campanelli2011cosmic,gruppuso2007complete,cea2014ellipsoidal}, rotating Universe \citep{godel1949example,ozsvath1962finite,ozsvath2001approaches,sivaram2012primordial,chechin2016rotation,seshavatharam2020integrated,camp2021}, or black hole cosmology \citep{pathria1972universe,easson2001universe,seshavatharam2014understanding,poplawski2010radial,tatum2018flat,christillin2014machian,seshavatharam2020light,chakrabarty2020toy}.

As discussed in \citep{shamir2020asymmetry,shamir2022possible2}, the observed anisotropy in the distribution of galaxy spin directions might also be driven by internal structure of galaxies rather than the large-scale structure of the Universe. In that case, the rotational velocity of the Milky Way relative to the rotational velocity of the observed galaxies would exhibit parity violation, forming an axis that is expected to peak at around the Galactic pole \citep{shamir2020asymmetry,shamir2022possible2,shamir2023dark,shamir2017large}. More information about the possible link between the anisotropy in galaxy spin directions and internal structure of galaxies is provided in \citep{shamir2022possible2}.

\section{Summary of previous work on asymmetry in the spin direction distribution of SDSS galaxies}
\label{previous_work}

Early analyses included a small number galaxies, suggesting the possibility of a non-random distribution of the spin directions of spiral galaxies  \citep{macgillivray1985anisotropy}. Analysis with a higher number of more than 6$\cdot10^3$ galaxies from the Southern hemisphere showed that the distribution is random \citep{iye1991catalog}. As explained in \citep{shamir2022using}, given the expected magnitude of the parity violation, the number of galaxies used in that study was too small to show a statistically significant parity violation. On the other hand, analysis of a larger number of galaxies from SDSS showed evidence of parity violation that forms a statistically significant dipole axis \citep{longo2007cosmic,longo2011detection}. In \citep{longo2011detection}, $\sim1.5\cdot10^4$ galaxies annotated by five undergraduate students were used to show non-randomness and a dipole axis in galaxy spin directions with statistical significance of $\sim5\sigma$. 

Another attempt to profile the large-scale distribution of galaxy spin directions was Galaxy Zoo 1, where SDSS galaxy images were annotated manually by anonymous volunteers through a web-based user interface \citep{land2008galaxy}. The results showed that according to the manual annotation, galaxies that spin counterclockwise are far more prevalent in SDSS compared to galaxies that spin clockwise. That large difference of $\sim$15\% was assumed to be the result of bias of the human perception or the user interface, rather than a reflection of the real distribution of spiral galaxies in the sky \citep{land2008galaxy}.

When the bias was noticed, a smaller set of galaxies was annotated again, but in that experiment the galaxies were also annotated after mirroring the images. Annotating both the original images and the mirrored images ensured that the annotation bias of the original images was offset by the annotation of the mirrored image. That experiment showed that indeed the large difference was driven by a certain bias in the annotation. After mirroring the images, 6.032\% of the galaxies were annotated as spinning counterclockwise, compared to 5.942\% of the mirrored galaxy images that were annotated as counterclockwise. Similarly, 5.525\% of the original galaxies images were annotated as spinning clockwise, compared to 5.646\% of the mirrored galaxy images that were annotated as spinning clockwise. These numbers are specified in Table 2 in \citep{land2008galaxy}.

In both cases, the number of galaxies spinning counterclockwise was $\sim$1.5\% or $\sim$2\% higher than the number of galaxies spinning clockwise. That difference agrees in both direction and magnitude with the asymmetry reported in \citep{shamir2020patterns}, which also used SDSS galaxies with spectra. Because just a small number of the galaxies were mirrored, the dataset contained just $\sim1.1\cdot10^4$ galaxies. The binomial statistical significance of the distribution was (P$\sim$0.13) when the clockwise galaxies were mirrored, and (P$\sim$0.21) when the counterclockwise galaxies were mirrored. The number of galaxies are shown in Table~\ref{entire_dataset}. These probabilities are not considered statistically significant, which can possibly result from the low number of galaxies, but the direction and magnitude of the distribution also does not conflict with the observed distribution of SDSS galaxies with spectra reported in \citep{shamir2020patterns}.

Another study proposed that the non-random distribution of galaxy spin directions in SDSS is the result of ``duplicate objects" in the data \citep{iye2020spin}. That study, however, does not refer to a specific paper that claimed for the presence of a dipole axis formed by the distribution of galaxy spin directions, and also had duplicate objects in the data. Also, a simple analysis showed that the ``clean'' data used in \citep{iye2020spin} is in fact not random \citep{shamir2022using}. Code and data to reproduce the analysis are available at \url{https://people.cs.ksu.edu/~lshamir/data/iye_et_al}. Clearly, the data used in the analysis contains no ``duplicate objects'', and shows that even after all duplicate objects are removed the distribution is not random. The statistical strength of a dipole axis in that specific dataset is $>2\sigma$, and therefore agrees with previous experiments that showed a dipole axis exhibited by the large-scale distribution of galaxy spin directions. More information about experiments and analysis of that dataset is provided in \citep{shamir2022using}.

Another analysis that examined the spin directions of galaxies with spectra in SDSS used the {\it SpArcFiRe} method to annotate a large number of SDSS galaxies \citep{hayes2017nature}. That dataset included the original Galaxy Zoo 1 galaxies \citep{lintott2008galaxy}. {\it SpArcFiRe} is a method that works best when applied to spiral galaxies, and therefore a first step of selecting spiral galaxies was applied before the galaxies were annotated by their spin direction using {\it SpArcFiRe}. The selection of spiral galaxies was done by two different methods. The first method was based on the manual annotation of the Galaxy Zoo volunteers, who annotated each galaxy as elliptical or spiral. After selecting the galaxies annotated as spiral and applying {\it SpArcFiRe} to identify their spin directions, the asymmetry between the number of clockwise and counterclockwise galaxies was statistically significant, and ranged between 2$\sigma$ to 3$\sigma$ \citep{hayes2017nature}. That led to the conclusion that the selection of spiral galaxies by Galaxy Zoo volunteers was biased in the sense that a galaxy that spins counterclockwise had a better chance of being labeled as spiral compared to a galaxy spinning clockwise. That was a new bias that was not noticed in the initial study of spin direction distribution in Galaxy Zoo \citep{land2008galaxy}.

To avoid the effect of a possible bias in the human perception, another analysis was performed by selecting the spiral galaxies by applying a machine learning classifier. The two-way classifier was trained with elliptical and spiral galaxies, and the class of spiral galaxies contained an equal number of clockwise and counterclockwise galaxies. That is, the number of galaxies spinning clockwise in the training set was exactly the same as the number of galaxies spinning counterclockwise. The equal number of spin directions in the class of spiral galaxies ensured that no certain spin direction would have a preference over the other spin direction in the selection of spiral galaxies. Clearly, that is a sound experimental design that ensured that no bias can result from arbitrary selection of a small set of galaxies in the training set.

But in addition to that careful design of the machine learning system, the machine learning algorithm was applied after identifying and removing the features that can identify to a certain level the spin direction of the galaxy. A stated in the paper, ``We choose our attributes to include some photometric attributes that were disjoint with those that Shamir (2016) found to be correlated with chirality, in addition to several SPARCFIRE outputs with all chirality information removed'' \citep{hayes2017nature}. While that implementation decision led to a random distribution of the annotated spin directions, it is also not clear whether the selection of the spiral galaxies was biased due to the removal of features that correlate with the spin direction, as it is expected that the removal of these features would lead to an even distribution of the annotations \citep{shamir2022using}. That is, the differences between galaxies with opposite spin directions could also be the source of an astronomical reason, rather than a bias in the algorithm. As shown in \citep{shamir2022using}, removing features that correlate with the spin direction can reduce the signal of the asymmetry. 

Here we perform a similar experiment and use the {\it SpArcFiRe} (SPiral ARC FInder and REporter) method to annotate the same set of Galaxy Zoo 1 galaxies, but by selecting spiral galaxies in three different manner: By manual analysis of Galaxy Zoo volunteers, by computer analysis, and with no selection of spiral galaxies at all. The selection of spiral galaxies is performed with no a-priori assumptions regarding their expected distribution. The data and code are available publicly to allow replication of the results. The possible scientific implications and the agreement of the observation with several other recent studies that make use of other probes \citep{aluri2022observable,luongo2022larger,dainotti2022evolution,yeung2022directional,hou2022measurement,philcox2022probing} are discussed in Section~\ref{conclusion}. Data for the experiments performed in this paper is made publicly available to allow reproduction of the results and further related experiments.

\section{Data}
\label{data}

\begin{figure}
\centering
\includegraphics[scale=0.85]{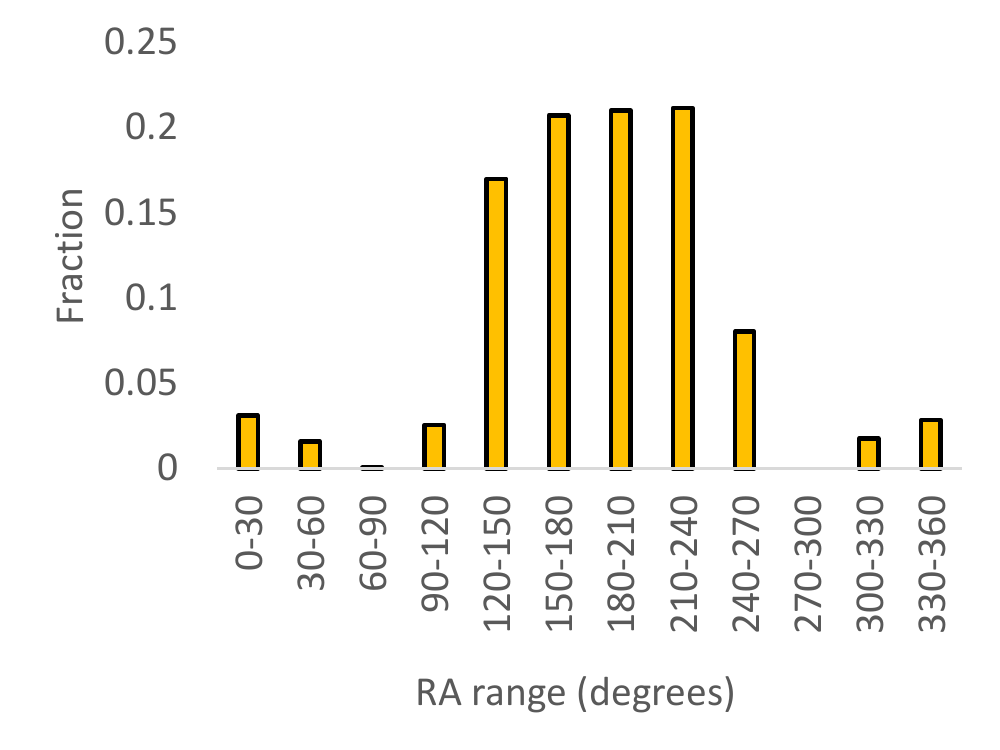}
\caption{The distribution of the galaxies in the dataset in different parts of the sky. The y-axis shows the fraction of the galaxies in each 30 degrees RA range out of the total number of galaxies in the dataset.}
\label{ra_distribution}
\end{figure}

The galaxies used in this study are SDSS galaxies used in Galaxy Zoo 1 \citep{lintott2008galaxy}. Images of 666,416 galaxies were downloaded in the JPEG file format using SDSS {\it cutout} service, and were converted to PNG for applying the {\it SpArcFiRe} (Scalable Automated Detection of Spiral Galaxy Arm) method \citep{Davis_2014,hayes2017nature}. The source code of {\it SpArcFiRe} is publicly available \footnote{https://github.com/waynebhayes/SpArcFiRe}. Annotation of a single 128$\times$128 galaxy image requires about 30 seconds of processing time when using a single Intel Core-i7 processor, and therefore the analysis was done by using 100 cores to reduce the response time of the analysis. Figure~\ref{ra_distribution} shows the distribution of the data into 30$^o$ RA bins. As the figure shows, the distribution of the galaxies in the sky is not uniform. That makes the analysis somewhat limited as the footprint of galaxies with spectra is practically smaller than what SDSS can provide, but the dataset is used for the sake of consistency and comparison with the previous studies that also used Galaxy Zoo 1 galaxies or SDSS galaxies with spectra. 

{\it SpArcFiRe} provides a detailed list of descriptors for each galaxy \citep{Davis_2014}. The method identifies arm segments in the image, and can group the pixels that are part of each segment. That allows to fit the pixels in the segment to a logarithmic spiral arc, which allows to extract different descriptors. For the spin direction, {\it SpArcFiRe} extracts several indicators, which are the longest arc, the majority of the arcs, the length weighted, and the pitch angle sum. The way the galaxy images are analyzed are explained thoroughly in \citep{Davis_2014}.

% https://academic.oup.com/mnras/article/466/4/3928/2733848
% https://iopscience.iop.org/article/10.1088/0004-637X/790/2/87

For the analysis we used galaxies for which all four indicators provided by {\it SpArcFiRe} and identify the spin direction of the galaxy showed the same spin direction. That provided a set of 273,055 galaxies with an annotated spin direction. The {\it SpArcFiRe} method was then applied again after mirroring the galaxy images, providing a set of 273,346 galaxies. After removing objects that were within 0.01 degrees or less to each other, the datasets were reduced to 271,063 and 271,308 galaxies, respectively. The slight differences between the results after mirroring the images is mentioned in \citep{hayes2017nature}, and will be discussed later in this paper.

To test the consistency of {\it SpArcFiRe}, we examined manually 322 galaxies and tested whether the annotation made by {\it SpArcFiRe} is in agreement with manual annotation. For that purpose, we identified manually 173 galaxies that by visual inspection seem to spin clockwise, and 149 galaxies that spin counterclockwise. From the clockwise galaxies, 122 galaxies were identified correctly as galaxies that spin clockwise, and 26 (15.02\%) as galaxies that spin counterclockwise. The rest of the galaxies were not annotated with a spin direction. Among the galaxies that were visually identified as galaxies spinning counterclockwise, 109 were also annotated by {\it SpArcFiRe} as counterclockwise, and 24 (16.1\%) as clockwise. The impact of the error will be discussed and analyzed in Section~\ref{results}.

Since {\it SpArcFiRe} is designed to analyze spiral galaxies, we performed a selection of just spiral galaxies in three manners: The first was selecting spiral galaxies that were annotated as spirals by the manual inspection of the Galaxy Zoo 1 volunteers \citep{lintott2008galaxy}. In Galaxy Zoo, each galaxy was annotated by several different annotators, who very often disagree with each other. To determine the annotation of a galaxy, a threshold is determined for the agreement between the different annotations. When the threshold is higher, the annotations are expected to be more accurate, but that also reduces the size of the dataset since less galaxies meet the higher threshold \citep{lintott2008galaxy}. Following \citep{hayes2017nature}, several experiments were made by selecting several different ``debiased'' thresholds.

Since the human selection of spiral galaxies can be biased, another method of selecting spiral galaxies was based on computer analysis. That was done by using the {\it Ganalyzer} method \citep{shamir2011ganalyzer}. As a model-driven method, it is not based on any kind of machine learning, and therefore it is not subjected to possible biases in the training data. The simple ``mechanical" nature of {\it Ganalyzer} allows it to be fully symmetric \citep{shamir2021large,shamir2022asymmetry}.

In addition to the manual selection and computer selection of spiral galaxies, another experiment was performed by using all galaxies that {\it SpArcFiRe} determined their spin direction without a first step of selection of spiral galaxies. While the annotation of galaxies that are elliptical can add noise to the system, it might be expected that the error in the annotation will be distributed equally between clockwise and counterclockwise galaxies. {\it SpArcFiRe} also does not force a certain spin direction, and can also annotate galaxies as not spinning in any identifiable direction. The list of galaxies and their annotations as assigned by {\it SpArcFiRe} is available at \url{https://people.cs.ksu.edu/~lshamir/data/sparcfire/}.

\section{Results}
\label{results}

A first experiment was a simple test of the distribution of spin directions in the entire dataset, and without any selection of spiral galaxies before applying {\it SpArcFiRe}. Another experiment was performed by selecting spiral galaxies by different thresholds of agreement of the Galaxy Zoo annotations, and then applying {\it SpArcFiRe} to annotate their spin direction. The selection of spiral galaxies was done by using different levels of agreement as thresholds of the Galaxy Zoo annotations. The agreement levels were between 40\% to 95\%. For instance, using 95\% as the agreement threshold means that only galaxies annotated as spiral by at least 95\% of the human annotators were selected. That also includes the ``clean" Galaxy Zoo standard of 80\%, and the ``superclean" \citep{lintott2008galaxy} standard of 95\%. Another method of selecting the spiral galaxies was by using the {\it Ganalyzer} model-driven algorithm for classifying between elliptical and spiral galaxies as described in Section~\ref{data}. The distributions of the spin directions in the entire dataset are shown in Table~\ref{entire_dataset}. Table~\ref{entire_dataset_previous} shows comparisons to previous studies using SDSS galaxies with spectra.

\begin{table}[h]
\caption{The distribution of galaxies spinning clockwise and counterclockwise in the entire dataset. The P values are the one-tailed binomial distribution probabilities when assuming 0.5 probability of a galaxy to spin clockwise or counterclockwise.}
\label{entire_dataset}
\scriptsize
\begin{tabular}{lcccc}
\hline
Spiral galaxy      & \# cw                  & \# ccw                  &  $\frac{\#ccw}{\#cw}$  & P  \\           
selection           &                            &                             &                        &   \\  
\hline
None           & 135,166 & 135,897 & 1.005 & 0.079 \\      % checked
None (mirrored) & 137,641 & 133,667 & 0.972 & $<10^{-5}$ \\
GZ1 40\%                & 84,812  & 85,017  & 1.002   &  0.308 \\ 
GZ1 40\% (mirrored)     & 85,772  & 84,273  & 0.982  & 0.0001 \\
GZ1 50\%                & 76,330  & 76,461  & 1.002   & 0.367  \\ 
GZ1 50\% (mirrored)     & 75,747  & 77,162  & 0.998  & 0.0001 \\
GZ1 80\%                & 44,132  & 44,733  & 1.014   & 0.022  \\ 
GZ1 80\% (mirrored)     & 43,877  & 44,983  & 0.986  & 0.0001 \\
GZ1 95\%                & 16,861  & 17,305  & 1.026   & 0.008  \\ 
GZ1 95\% (mirrored)     & 16,850  & 17,339  & 0.972 & 0.004 \\
Computer                         & 69,043   & 69,897 & 1.012 & 0.01 \\                            % checked
Computer (mirrored)         & 70,743  & 69,180  & 0.977 & 1.5$\cdot10^{-5}$ \\   % checked
\hline
\end{tabular}
\end{table}

\begin{table}[h]
\caption{The distribution of galaxies spinning clockwise and counterclockwise in previous reports using SDSS galaxies with spectra.}
\label{entire_dataset_previous}
\scriptsize
\begin{tabular}{lcccc}
\hline
Report      & \# cw                  & \# ccw                  &  $\frac{\#ccw}{\#cw}$  & P  \\           
\hline
\citep{land2008galaxy}  (ccw mirrored)  & 5,044 & 5,133   & 1.018 & 0.18 \\   % from a total of 91,303
\citep{land2008galaxy}  (cw mirrored)         & 5,425 & 5,507  & 1.015 & 0.21 \\
\citep{longo2011detection}                                  & 7,442 & 7,816   & 1.05  & 0.001 \\
\citep{shamir2020patterns}                                  & 32,055 & 32,501 & 1.014 & 0.039 \\
\hline
\end{tabular}
\end{table}

As the table shows, all experiments show a higher number of galaxies spinning counterclockwise than clockwise. When mirroring the galaxy images, {\it SpArcFiRe} shows a higher number of clockwise galaxies, which are in fact galaxies spinning counterclockwise in the original, non-mirrored, galaxy images. When mirroring the images, the results are not completely inverse to the results when using the original images. That is not surprising, since it has been reported that {\it SpArcFiRe} has certain degree of asymmetry in the manner it annotates galaxy images. As explained in Appendix A of \citep{hayes2017nature}, {\it SpArcFiRe} is not fully symmetric, and therefore the galaxies were annotated again after mirroring the images. The complexity of {\it SpArcFiRe} made it difficult to identify the reasons for the differences between the original and mirrored images \citep{hayes2017nature}.

The results are also compared to previous literature of experiments that used SDSS galaxies with spectra, and were based on different annotation methods. These experiments use symmetric automatic annotation \citep{shamir2020patterns} or manual annotation such that the galaxy images were mirrored \citep{land2008galaxy,longo2011detection}. All of these experiments also show a higher number of galaxies spinning counterclockwise. This agreement does not necessarily prove that the observed asymmetry is not driven by a combination of bias and statistical fluctuations, but they also do not conflict with each other. The only inconsistency is the far greater asymmetry of $\sim$5\% observed by \citep{longo2011detection}, which is far higher than the asymmetry reported by all other studies. That includes Galaxy Zoo, which also annotates the galaxies from the same footprint of SDSS galaxies with spectra by using manual annotation. 

The asymmetry shown when the spirals galaxies are selected manually by Galaxy Zoo volunteers is in agreement with the results shown in \citep{hayes2017nature}. For the automatic selection of the galaxies, the results disagree with \citep{hayes2017nature}. A possible reason for that disagreement is that the automatic selection of spiral galaxies performed in \citep{hayes2017nature} were done after applying a machine learning algorithm designed by  specifically removing the attributes that correlate with the galaxy spin direction. The spiral selection used for the results shown in Table~\ref{entire_dataset} are done by a simple symmetric model-driven algorithm, and therefore without selection or removal of specific attributes.

\subsection{Identification of a possible dipole axis alignment}
\label{dipole}

Previous work using different telescopes showed that the spin directions of spiral galaxies form a statistically significant large-scale axis \citep{shamir2022analysis}. That was done by fitting the spin directions to the cosine of the angle between the galaxies and every possible integer $(\alpha,\delta)$ combination in the sky \citep{shamir2022asymmetry}, as shown by Equation~\ref{chi2}
\begin{equation}
\chi^2_{(\alpha,\delta)}=\Sigma_i | \frac{(d_i \cdot | \cos(\phi_i)| - \cos(\phi_i))^2}{\cos(\phi_i)} | ,
\label{chi2}
\end{equation}
where $d_i$ is 1 if galaxy {\it i} spins clockwise or -1 if galaxy {\it i} spins counterclockwise, and $\phi_i$ is the angular distance between galaxy {\it i} and the location of the possible dipole axis $(\alpha,\delta)$.

\begin{figure*}[h]
\centering
\includegraphics[scale=0.25]{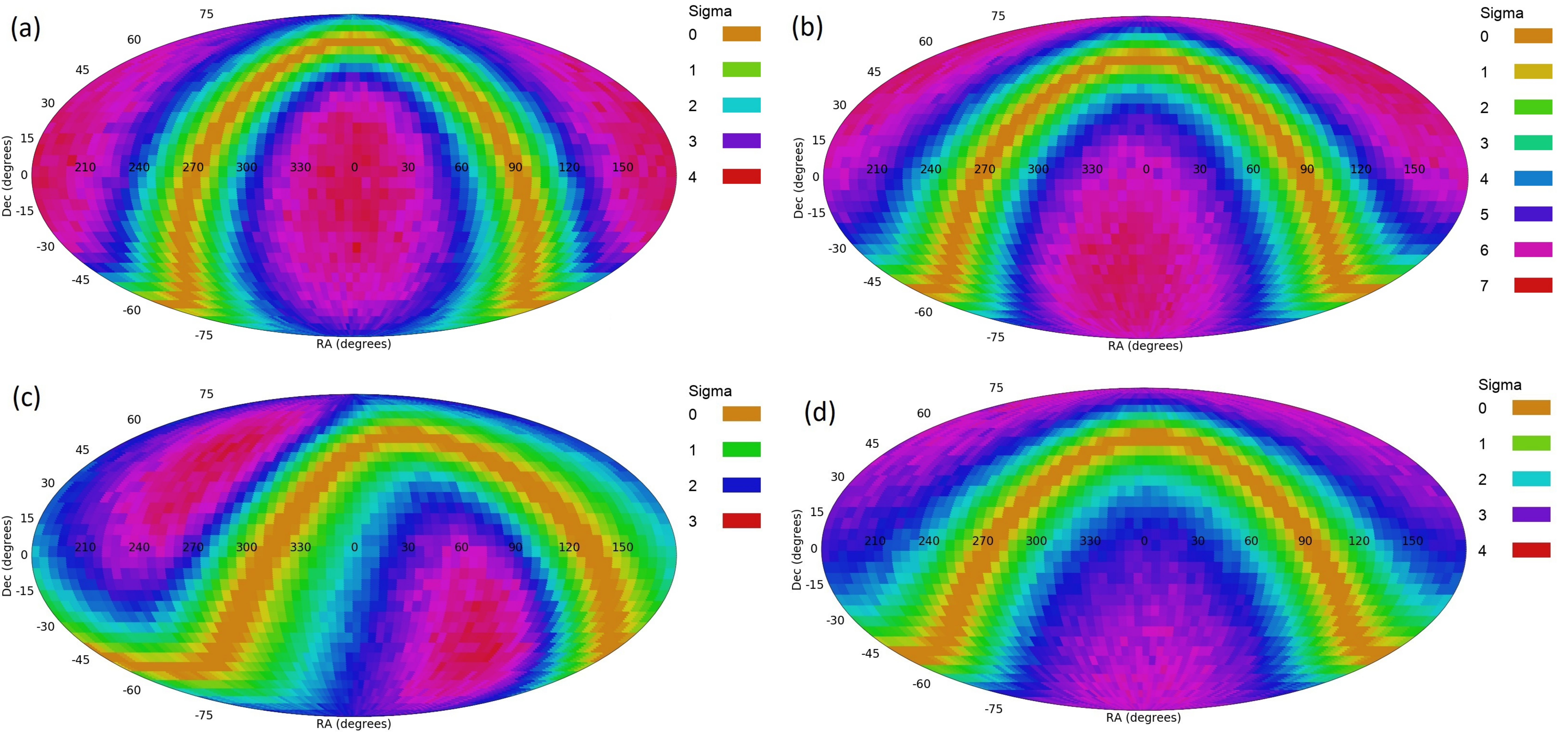}
\caption{The $\chi^2$ statistical significance of a dipole axis from different $(\alpha,\delta)$ when using {\it SpArcFiRe} for the annotation of the galaxy spin directions and different methods of selecting spiral galaxies: a) computer selection, b) no spiral selection, c) Galaxy Zoo selection with 95\% threshold (``superclean"), and d) Galaxy Zoo selection with 80\% threshold (``clean").
}
\label{dipole_all}
\end{figure*}

The statistical significance of the possible dipole axis centered at $(\alpha,\delta)$ is determined by assigning the galaxies with random spin directions, and computing the $\chi^2_{(\alpha,\delta)}$ using Equation~\ref{chi2}. That is done 1000 times, and the mean and $\sigma$ of the $\chi^2_{(\alpha,\delta)}$ are determined. The $\sigma$ difference between the mean when $\chi^2_{(\alpha,\delta)}$ is computed by random spin directions and the $\chi^2_{(\alpha,\delta)}$ computed when using the observed spin directions determines the statistical strength of the axis. Repeating that process from each possible $(\alpha,\delta)$ integer combination shows the statistical signal of a dipole axis at all parts of the sky. Figure~\ref{dipole_all} shows several examples of applying the analysis to the datasets in Table~\ref{entire_dataset} that had the lowest P values.

As the figure shows, the profiles exhibited by the different methods of selecting spiral galaxies are similar to each other. The automatic selection shows results in agreement with the manual selection of spiral galaxies, but the statistical significance is different. Table~\ref{dipole_axes} shows the locations of the most likely dipole axis when using the different methods of spiral galaxy selection. The locations of the dipole axis are somewhat different across different datasets, but in all cases still within 1$\sigma$ difference compared to each other.

\begin{table*}
\caption{Most likely locations of the dipole axes observed in the different datasets.}
\label{dipole_axes}
\scriptsize
\begin{tabular}{lccccc}
\hline
Dataset      & RA ($^o$)       & Dec ($^o$)       &  $\sigma$ & RA error ($^o$) & Dec error ($^o$)\\           
\hline
No spiral selection            & 129 & 14 & 1.91 & 29-258  & -63-90 \\
No spiral selection (mirrored) & 170 & 35 & 6.88  & 77-230  & -12-90 \\
Galaxy Zoo 40\%                & 268  & 27  &  1.89  & 151-349  & -44-85 \\ 
Galaxy Zoo 40\% (mirrored)     & 147  & 22  & 3.67  &  50-224 & -44-90 \\
Galaxy Zoo 50\%                & 243  &  14 &  1.19  & 122-345  &  -78-90  \\ 
Galaxy Zoo 50\% (mirrored)     &  149 & 19  &  3.59  & 51-231 & -42-90 \\
Galaxy Zoo 80\%                & 184  & 52  &  1.81   & 69-282  & -32-90  \\ 
Galaxy Zoo 80\% (mirrored)     & 177  & 41  &  3.57 & 66-275  & -90-34 \\
Galaxy Zoo 95\%                & 146 & 31  &  2.84   & 61-214  & -37-90 \\ 
Galaxy Zoo 95\% (mirrored)     & 244 & 28  &  2.94  & 171-291  & -36-90 \\
Computer                            & 192  & 24 &  2.33  &  123-253 & -69-90  \\  
Computer (mirrored)            & 184  & 16  &  3.97  & 128-238  & -35-90 \\
\hline
\end{tabular}
\end{table*}

An interesting observation is that the statistical significance of the axis is stronger when the galaxy images were mirrored. That agrees with the simple statistical significance shown in Table~\ref{entire_dataset}. That can be the result of a certain asymmetric behavior of the annotation algorithm in the case that the distribution of spin directions in the sky is random. If the distribution of galaxy spin directions is not random, the difference can be explained by a certain bias of the algorithm. That is discussed in detail later in this section.

\begin{figure*}[h]
\centering
\includegraphics[scale=0.25]{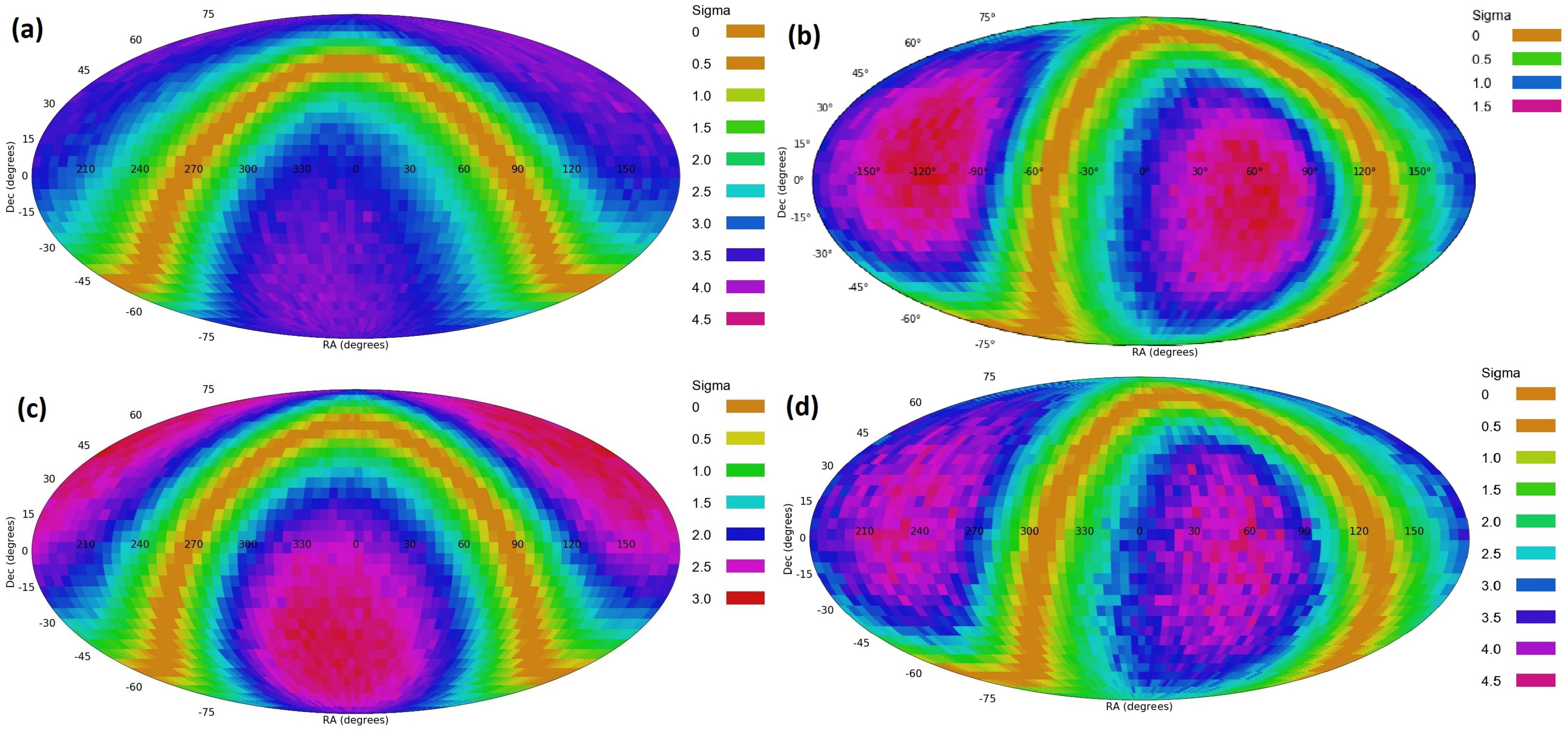}
\caption{Previous results of dipole axes observed in previous work. (a) is Figure 4 in \citep{shamir2021particles} based on 77,840 SDSS galaxies \citep{shamir2021particles}, (b) is Figure 12 in \citep{shamir2020patterns} based on Pan-STARRS data, (c) is Figure 13 in \citep{shamir2022analysis2} based on SDSS data, and (d) is Figure 3 in \citep{shamir2021large} based data from DECam. }
\label{dipole_previous_square}
\end{figure*}

The axes can also be compared to previous experiments with 77,840 SDSS galaxies \citep{shamir2021particles}. That dataset contains galaxies that do not necessarily have spectra, but these galaxies are relatively large (Petrosian radius $>$ 5.5') and bright (i magnitude $<$ 18). Figure~\ref{dipole_previous_square} shows the results of previous experiments \citep{shamir2021particles} when using SDSS galaxies that do not necessarily have spectra, as well as another dataset of 13,440 SDSS galaxies with spectra originally used in \citep{shamir2016asymmetry}. In the experiment of \citep{shamir2021particles} the galaxies were annotated automatically by using a model-driven symmetric annotation method. The galaxies in \citep{shamir2016asymmetry} were annotated with manual inspection.

Other experiments that can be used for comparison include other telescopes. These include an experiment with 33,028 galaxies imaged by Pan-STARRS \citep{shamir2020patterns} and 807,898 galaxies imaged by DECam \citep{shamir2021large}. Figure~\ref{dipole_previous_square} shows that in these experiments the results are similar to the results shown in Figure~\ref{dipole_all}. % The datasets used in this experiment are both publicly available. The dataset used in \citep{shamir2021particles} is available at \url{https://people.cs.ksu.edu/~lshamir/data/assymdup}, and the dataset used in \citep{shamir2022analysis2} is available at \url{https://people.cs.ksu.edu/~lshamir/data/assym}. The most likely dipole axis observed in Pan-STARRS is $(\alpha=197^o,\delta=2^o)$, and the 

The experiment of \citep{longo2011detection} showed a dipole axis that peaks at $(\alpha=217^o,\delta=32^o)$. That location is also within 1$\sigma$ statistical error to the most likely axes shown here, as specified in Table~\ref{dipole_axes}. It is also close to the location of the dipole axis observed with Pan-STARRS \citep{shamir2020patterns} at $(\alpha=197^o,\delta=2^o)$, and DECam \citep{shamir2021large} at $(\alpha=237^o,\delta=10^o)$.

\begin{figure}[h]
\centering
\includegraphics[scale=0.66]{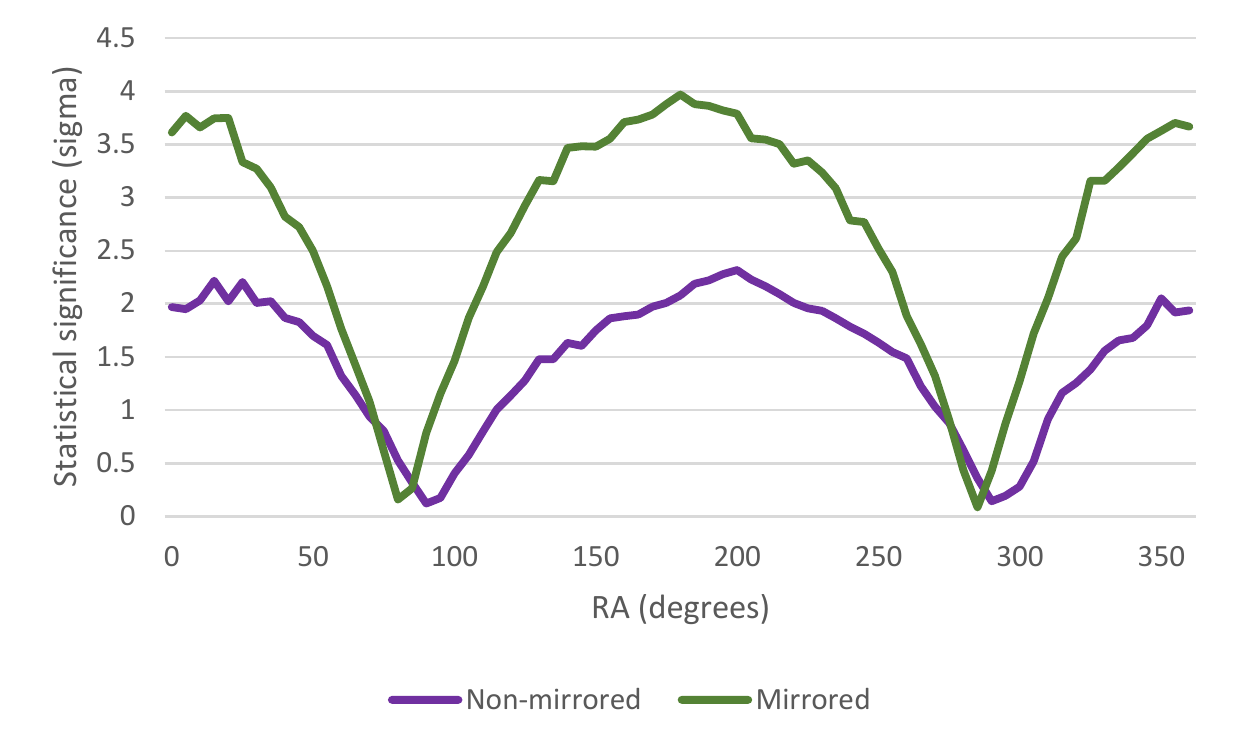}
\caption{$\chi^2$ statistical significance of cosine dependence of the galaxy spin directions from different RAs when the declination is 25$^o$. The analysis was done with the original and mirrored galaxy images annotated by {\it SpArcFiRe}. }
\label{dec25}
\end{figure}

When using the automatically selected spiral galaxies, the statistical significance of the dipole axis is maximal at around declination of $25^o$. Figure~\ref{dec25} displays the $\chi^2$ statistical significance at different RAs when the declination is set to $25^o$. As the figure shows, the analysis with the mirrored galaxy images and the original galaxy images show similar profiles, but the statistical signal is significantly stronger when the mirrored galaxy images were used. That shows that the asymmetry of {\it SpArcFiRe} as reported in \citep{hayes2017nature} can affect the statistical signal. That is also shown in Table~\ref{entire_dataset}.

\begin{figure}
\centering
\includegraphics[scale=0.66]{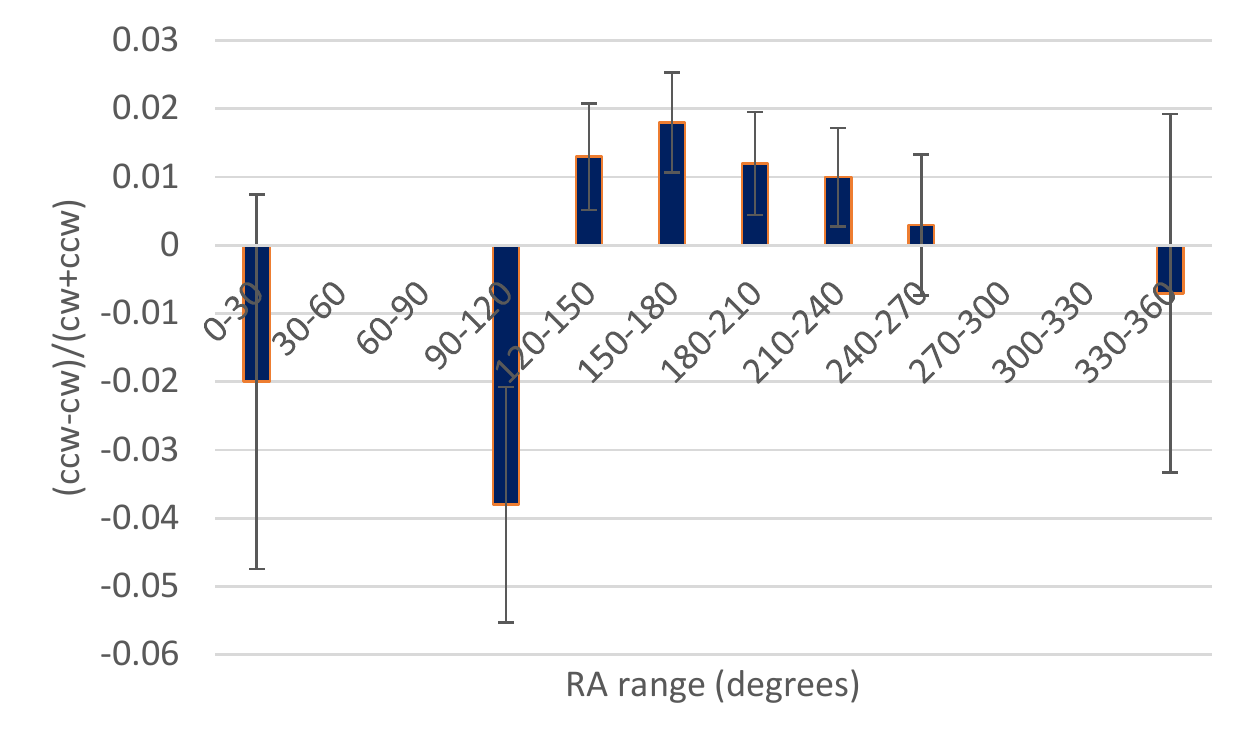}
\caption{Asymmetry in different RA ranges.}
\label{ra_ranges}
\end{figure}

Figure~\ref{ra_ranges} shows a simple analysis of the simple asymmetry in different RA ranges, when the declination range is 5$^o$ to 45$^o$. The figure shows a higher number of galaxies spinning counterclockwise in the RA range of $(120^o,270^o)$, and the asymmetry peaks at $(150^o,180^o)$. In the other hemisphere there are more galaxies spinning clockwise, but due to the small total number of galaxies in that hemisphere it is difficult to profile that asymmetry.

\subsection{Analysis of possible algorithm bias}
\label{bias_analysis}

One of the explanations for the results shown here is a bias in the {\it SpArcFiRe} annotation algorithm, and it has been reported that such subtle asymmetry exists \citep{hayes2017nature}. For instance, if {\it SpArcFiRe} tends to prefer to annotate galaxies as spinning counterclockwise, such consistent bias can become statistically significant. It has been shown that such bias can also lead to a dipole axis. It has been shown that even a subtle but consistent bias in the annotation algorithm can lead to a dipole axis with extremely high statistical signal, that peaks exactly at the celestial pole \citep{shamir2021particles}. 

One of the experiments done to study the nature of the bias is repeating the experiments after mirroring the galaxy images. Based on the results shown here, if the algorithm is systematically biased to prefer a certain spin direction, it would have been a preference to galaxies that spin counterclockwise. That is, the results  shown here can be explained by a bias in the algorithm, or by a bias in the selection of spiral galaxies.

Assuming an equal number of clockwise and counterclockwise galaxies in the sky and a bias {\it b} in the {\it SpArcFiRe} software, the number of clockwise galaxies will be lower than the number of galaxies annotated as counterclockwise, which is $(1+b)ccw$. The asymmetry {\it A} between the number of clockwise and counterclockwise galaxies can be defined as $A=\frac{(1+b) \cdot ccw}{cw}$, where {\it cw} is the number of galaxies that spin clockwise {\it ccw} is the number of galaxies that spin counterclockwise, and $b$ is the bias such that $b>0$. When mirroring the galaxy images, {\it A} can be defined as $A_{mirrored}=\frac{(1+b) \cdot cw}{ccw}$. Assuming no asymmetry in the sky, cw is equal to ccw, and therefore $A_{mirrored}$ is equal to $\frac{(1+b) \cdot ccw}{cw}$, which is equal to $A$. That, however, is not what is observed when mirroring the galaxy images. Mirroring the galaxy images provides different results, and also flips the sign of the asymmetry as shown consistently in Table~\ref{entire_dataset}. 

On the other hand, assuming that the real sky distribution of spin directions of spiral galaxies is not symmetric, $ccw > cw$. The asymmetry ratio $a$ between the number of clockwise and counterclockwise galaxies can be defined as $a=\frac{ccw}{cw}$. The asymmetry $a-1$ is positive when the number of $cc$ galaxies is higher than the number if $cw$ galaxies, and negative if the number of $cw$ galaxies is larger. The asymmetry $A$ of the original non-mirrored images can be defined as 

\begin{equation*}
A=\frac{(1+b) \cdot ccw}{cw}=\frac{(1+b) \cdot a \cdot cw}{cw}=(1+b) \cdot a = a + ab .
\end{equation*}

{\it A} can be greater or smaller than 1, depends on the values of $a$ and $b$. After mirroring the galaxies, the asymmetry of the mirrored galaxies $A_{mirrored}$ is

\begin{equation*}
A_{mirrored}=\frac{(1+b) \cdot cw}{ccw} = \frac{(1+b) \cdot cw}{a \cdot cw} =  \frac{1+b}{a}  .  %  =\frac{1}{a}+\frac{b}{a} 
\end{equation*}

If $ccw>cw$, {\it a} is greater than 1. if $b$ is positive $A$ will necessarily be greater than 1, which is the observation shown in Table~\ref{dipole_axes} for the non-mirrored images. $A_{mirrored}$ can be either greater or smaller than 1. If $b>a-1$, $A_{mirrored}$ will be greater than 1, otherwise $A_{mirrored}$ will be smaller than 1. Since the observed $A_{mirrored}$ is in all cases smaller than 1, the asymmetry of the algorithm $b$ is smaller than the asymmetry $a-1$ of the spin directions of the galaxies in the dataset. While this simple analysis is expected, it shows that if the $A$ drops from a number greater than 1 when using the original images to a number smaller than 1 when using the mirrored images, that change is not driven by the asymmetry of the algorithm.

Another possible reason for the observed results can be a bias in the selection of spiral galaxies. A selection of a spiral galaxy is not necessarily a formally defined task, and the separation between spiral and elliptical galaxies have many in-between cases. If more counterclockwise galaxies are selected as spiral galaxies compared to clockwise galaxies, that will result in a dataset that has a higher number of counterclockwise galaxies. 

In previous work the bias was addressed by comparing the distribution in two opposite hemispheres \citep{shamir2020patterns,shamir2021large,shamir2021particles,shamir2022asymmetry,shamir2022analysis}. In these experiments, the galaxies were separated into two hemispheres such that one hemisphere showed a higher number of galaxies spinning clockwise, and the opposite hemisphere showed a higher number of galaxies spinning counterclockwise. Table~\ref{hemispheres} shows the distribution of 807,898 galaxies imaged by DECam, as thoroughly described in \citep{shamir2021large}.

\begin{table*}
\caption{The number of galaxies spinning clockwise and galaxies spinning counterclockwise in opposite hemispheres. The table is taken from \citep{shamir2021large}.}
\label{hemispheres}
\centering
\begin{tabular}{lcccc}
\hline
Hemisphere       & \# cw galaxies & \# ccw galaxies  & $\frac{cw-ccw}{cw+ccw}$  & P \\
\hline
$(0^o-150^o \cup 330^o-360^o)$ &   264,707    &  262,559  &   0.004      &  0.0015  \\   
$(150^o-330^o)$                          &   139,719     & 140,913  &   -0.004    &  0.0121   \\
\hline
\end{tabular}
\end{table*}

If the selection of spiral galaxies was biased, it is expected that the selection would have been consistent in all parts of the sky. That is, if more counterclockwise galaxies are selected as spiral galaxies, that should lead to a higher number of counterclockwise galaxies in all parts of the sky, and is not expected to flip in opposite hemispheres of the sky. In the SDSS galaxies with spectra used in Galaxy Zoo 1, the vast majority of the galaxies are concentrated in one hemisphere, with very few galaxies in the opposite hemisphere as shown in Figure~\ref{ra_distribution}. The absence of galaxies in two opposite hemispheres makes it difficult to apply the same analysis as done in \citep{shamir2020patterns,shamir2021large,shamir2021particles,shamir2022asymmetry,shamir2022analysis}, and the example shown in Table~\ref{hemispheres}. 

An attempt to follow that analysis with the SDSS galaxies used in this study is shown in Tables~\ref{hemisphere1} and~\ref{hemisphere2}. Table~\ref{hemisphere1} shows the distribution of galaxy spin directions of the SDSS galaxies annotated by {\it SpArcFiRe}, such that the RA of the galaxies fall within $(120^o<\alpha<330^o)$. Table~\ref{hemisphere2} shows the same analysis for the opposite hemisphere $(\alpha<120^o \cup \alpha>330^o)$. The results show that in the more populated part of the sky the higher number of counterclockwise galaxies is consistent and statistically significant. In the opposite part of the sky, the number of clockwise galaxies is higher, but the statistical significance is low. That can be the result of the far lower number of galaxies in that part of the sky, not allowing a strong statistical signal.

\begin{table}[h]
\caption{The distribution of galaxy spin directions in the hemisphere $(120^o<\alpha<330^o)$. The analysis is done by applying the {\it SpArcFiRe} algorithm after selecting spiral galaxies using Galaxy Zoo (GZ) annotations, computer annotations, and by applying {\it SpArcFiRe} with no selection of spiral galaxies at all.}
\label{hemisphere1}
\scriptsize
\begin{tabular}{lcccc}
\hline
Spiral galaxy      & \# cw                  & \# ccw                  &  $\frac{\#ccw}{\#cw}$  & P  \\           
selection          &                        &                         &                        &   \\  
\hline
None                & 99,422 & 100,681 & 1.013 &  0.002 \\
None (mirrored) & 101,983  & 98,456 & 0.965 & $<10^{-5}$ \\
GZ 40\%                &  53,711 & 53,960 & 1.005   &  0.22 \\ 
GZ 40\% (mirrored)     &  54,448 &  53,385 & 0.98  & 0.0006 \\
GZ 50\%                & 46,056  & 46,326  &  1.006  &  0.19 \\ 
GZ 50\% (mirrored)     & 46,711  & 45,759  &  0.98 & 0.0009 \\
GZ 80\%                &  39,884 & 40,496  & 1.015   &  0.015 \\ 
GZ 80\% (mirrored)     &  40,766 & 39,662  & 0.973  & $<10^{-4}$ \\
GZ 95\%                & 15,327  &  15,742 &  1.027  & 0.01  \\ 
GZ 95\% (mirrored)     & 15,793  & 15,333  & 0.971  &  0.005 \\
Computer                       &  62,301 & 63,227 & 1.014 &  0.004 \\  
Computer (mirrored)            & 63,968 &  62,369 &  0.975 & $<10^{-5}$ \\
\hline
\end{tabular}
\end{table}

\begin{table}[h]
\caption{The distribution of galaxy spin directions in the hemisphere $(\alpha<120^o \cup \alpha>330^o)$.}
\label{hemisphere2}
\scriptsize
\begin{tabular}{lcccc}
\hline
Spiral galaxy      & \# cw                  & \# ccw                  &  $\frac{\#ccw}{\#cw}$  & P  \\           
selection          &                        &                         &                        &   \\  
\hline
None            & 9,884 & 9,781 & 0.99 & 0.23 \\
None (mirrored) & 9,747 & 9,898 & 1.015 & 0.13 \\
GZ 40\%                &  4,479 & 4,392 &  0.98  & 0.18  \\ 
GZ 40\% (mirrored)     & 4,418  &  4,442 &  1.005 & 0.39 \\
GZ 50\%                &  3,578 &  3,544 &  0.99  &  0.34 \\ 
GZ 50\% (mirrored)     & 3,545  & 3,542  &  0.999 & 0.49 \\
GZ 80\%                &  3,080 &  3,048 &  0.994  &  0.46 \\ 
GZ 80\% (mirrored)     &  3,066 & 3,048  & 0.994  & 0.41 \\
GZ 95\%                &  1,118 &  1,119 &  1.001  &  0.48 \\ 
GZ 95\% (mirrored)     & 1,101  & 1,108  & 1.006 & 0.43 \\
Computer                       & 7,112  & 7,001 & 0.984 &  0.18 \\  
Computer (mirrored)            & 7,033  &  7,111 & 1.011 & 0.25 \\
\hline
\end{tabular}
\end{table}

Table~\ref{hemisphere2} shows an experiment with a small number of galaxies, and does not allow to determine a statistically significant asymmetry. The table shows certain evidence of a higher number of galaxies that spin clockwise in that hemisphere, but the number of galaxies is not sufficient to determine a statistically significant opposite asymmetries in spin directions.

Because the distribution of the galaxies in the sky make the separation of the dataset into two hemispheres impractical, two other methods of selecting spiral galaxies were applied in addition to the manual selection of spiral galaxies by Galaxy Zoo volunteers. The first was to apply {\it SpArcFiRe} with no selection of spiral galaxies. When {\it SpArcFiRe} cannot identify the spin direction of the galaxy, the galaxy is not used. The disadvantage of that method is that without a first step of selecting spiral galaxies {\it SpArcFiRe} might provide less accurate annotations. The second method that was for selecting spiral galaxies was by using the {\it Ganalyzer} \citep{shamir2011ganalyzer} algorithm, which is a simple model-driven method that can identify spiral galaxies. {\it Ganalyzer} is far less sophisticated than {\it SpArcFiRe}, and provides less information about the galaxy. On the other hand, its simple ``mechanical'' nature allows it to be fully symmetric. The symmetric nature of {\it Ganalyzer} was tested in previous studies \citep{shamir2020patterns,shamir2021large,shamir2021particles,shamir2022asymmetry,shamir2022analysis}. When selecting spiral galaxies with {\it Ganalyzer}, the number of counterclockwise galaxies is higher when analyzing the original images, and the number of clockwise galaxies is higher when analyzing the mirrored images.

The analysis is challenged by the fact that {\it SpArcFiRe} is not fully symmetric. The observation that the sign of the asymmetry flips when the images are mirrored indicates that the asymmetry of {\it SpArcFiRe} is smaller than the asymmetry between the number of clockwise and counterclockwise galaxies in the set of SDSS galaxies with spectra. While the results shown in this paper might not be sufficient to prove non-random distribution of galaxy spin directions, they show that the distribution of the spin directions of SDSS galaxies with spectra are in agreement with non-random distribution, and do not conflict with previous results.

\section{Conclusion}
\label{conclusion}

The availability of large digital sky surveys enabled by high-throughput robotic telescopes has enabled the studying of questions that were not addressable in the pre-information era. The distribution of spin directions of spiral galaxies is a question that was studied by using several sky surveys and several analysis methods. The set of SDSS galaxies with spectra is one of the datasets that was studied several times in the past, showing different conclusions. One of these studies used the {\it SpArcFiRe} method to annotate SDSS galaxies with spectra used in Galaxy Zoo 1 \cite{hayes2017nature}.

The experiment performed here used the same {\it SpArcFiRe} method of annotation that was used in \citep{hayes2017nature}. While {\it SpArcFiRe} was used to annotate the spin directions of the spiral galaxies, that annotation was applied after a first step of selecting the spiral galaxies and separating them from the rest of the galaxies. When the spiral galaxies are selected manually by Galaxy Zoo volunteers, the number of clockwise and counterclockwise galaxies is not symmetric, as was also reported in \citep{hayes2017nature}. But when selecting the spiral galaxies automatically, or when not selecting the spiral galaxies at all, the number of galaxies spinning clockwise is also significantly different from the number of galaxies spinning counterclockwise.

\cite{hayes2017nature} also performed an experiment by selecting the spiral galaxies automatically, and used a machine learning algorithm for that task. The algorithm was trained with elliptical and spiral galaxies, such that the class of spiral galaxies contained an equal number of clockwise and counterclockwise galaxies. Such construction of the training set can avoid a situation in which more galaxies of a certain spin direction are classified as spiral. From a machine learning perspective, that is a careful design that aims at reducing the possible biases introduced by machine learning.

But in addition to selecting spiral galaxies, the machine learning algorithm was applied after manually removing all attributes that correlated with the spin direction. As shown in \citep{shamir2022using}, when using machine learning to select spiral galaxies, removing specifically the features that correlate with spin direction leads to random distribution of the spin directions. In this paper the experiments were performed by selecting the spiral galaxies without removing specific features, and in fact without using machine learning. The results show that the spin directions of spiral galaxies as seen from Earth form a dipole axis with statistical significance of between 2.33$\sigma$ to 3.97$\sigma$ .Some of the experiments were also done by applying {\it SpArcFiRe} without a first step of selection of spiral galaxies. In all cases the results are consistent, and also showed statistically significant dipole axis formed by the spin directions of the galaxies.

Due to the limited footprint size, the results of the SDSS data annotated by {\it SpArcFiRe} as shown here cannot provide the comprehensive analysis of a very large footprint such as the DESI Legacy Survey \citep{shamir2022analysis}. But the results shown here are in agreement with previous reports that use different telescopes and different analysis methods. These experiments include SDSS galaxies \citep{shamir2020patterns}, but also consistent across other telescopes such as Pan-STARRS \citep{shamir2020patterns}, DECam \citep{shamir2021large}, Hubble Space Telescope \citep{shamir2020pasa}, the Dark Energy Survey \citep{shamir2022asymmetry}, and the DESI Legacy Survey \citep{shamir2022analysis}. 

Because {\it SpArcFiRe} has a small but consistent asymmetry, using {\it SpArcFiRe} for this task is more difficult compared to fully symmetric methods, as discussed thoroughly in Section~\ref{bias_analysis}. Simple binomial distribution shows a maximum probability of 1.5$\cdot10^{-5}$ to occur by chance. When mirroring the images, the asymmetry is inverse, and the statistical signal is still significant at 0.01. That shows statistically significant parity violation in galaxy spin directions when using {\it SpArcFiRe} to annotate SDSS galaxies. 

Analysis of a dipole axis formed by the distribution of galaxy spin directions shows different levels of statistical significance depends on the selection of the galaxies and size of the dataset, and the experiments agree with the contention that a dipole axis exists with statistical significance as high as $\sim4\sigma$. None of the experiments showed disagreement with the presence of a dipole axis. An interesting observation is that the most likely position of the dipole axis is at close proximity to the galactic pole, which might indicate that the dipole axis is not related to the large-scale structure of the local Universe but to internal structure of galaxies. That is discussed in \citep{shamir2022possible2,shamir2023dark,shamir2023doppler}.

\section*{Data Availability}

Data used in this study were made available publicly. The list of SDSS galaxies and their spin direction annotations used in this study is available at \url{https://people.cs.ksu.edu/~lshamir/data/sparcfire/}. Annotations of the mirrored galaxies are also available at the same URL. Additionally, a list of galaxies and their spin directions used for creating panel (a) in Figure~\ref{dipole_previous_square} is available at \url{https://people.cs.ksu.edu/~lshamir/data/assymdup/}. A list of the smaller set of galaxies used to create Figure~\ref{dipole_previous_square} is available at \url{https://people.cs.ksu.edu/~lshamir/data/assym/}. The galaxies used in \citep{shamir2022using} and reproduction of that experiment are available at \url{https://people.cs.ksu.edu/~lshamir/data/iye_et_al}.

\section*{Conflict of Interest}
              
The authors declare that they have no conflicts of interest.

\section*{Acknowledgments}
              
We would like to thank the knowledgeable reviewer for the insightful comments. The research was funded in part by NSF grant AST-1903823.

\bibliographystyle{apalike}

\bibliography{main}

\end{document}